\documentstyle[jkas35]{article}

\runningauthor{AHN AND SHAPIRO}
\runningtitle{SELF-INTERACTING DARK MATTER HALOS}
\beginpage{1}
\endpage{11}

\begin{document}

\title{FORMATION AND EVOLUTION OF SELF-INTERACTING DARK MATTER HALOS}

\author{Kyungjin Ahn and Paul R. Shapiro}

\address{Department of Astronomy, University of Texas, 
Austin, TX 78712, USA \\
{\it E-mail: kjahn@astro.as.utexas.edu; shapiro@astro.as.utexas.edu}}

\address{\normalsize{\it (Received mmm, dd, 2002; Accepted ???. ??, 2002)}}

\abstract{
Observations of dark matter dominated dwarf and low surface brightness disk 
galaxies favor density profiles with a flat-density core, while cold dark 
matter (CDM) N-body simulations form halos with central cusps, instead.
This apparent discrepancy has motivated a re-examination of the microscopic 
nature of the dark matter in order to explain the observed halo profiles, 
including the suggestion that CDM has a non-gravitational self-interaction.
We study the formation and evolution of self-interacting dark matter
(SIDM) halos. We find analytical, fully
cosmological similarity solutions for their dynamics, which take proper 
account of the collisional interaction of SIDM particles, based on
a fluid approximation derived from the Boltzmann equation.
The SIDM particles scatter each other elastically, which results in an
effective thermal conductivity that heats the halo core and flattens its
density profile. 
These similarity solutions are relevant to galactic and cluster halo 
formation in the CDM model.
We assume that the local density maximum
which serves as the progenitor of the halo has an initial mass profile
\( \delta M/M\propto M^{-\varepsilon } \), as in the familiar secondary
infall model. If \( \varepsilon =1/6 \), SIDM halos will evolve self-similarly,
with a cold, supersonic infall which is terminated by a strong accretion
shock. Different solutions arise for different values of the dimensionless
collisionality 
parameter, \( Q\equiv\sigma \rho_{b}r_{s} \), where \( \sigma  \) is the
SIDM particle scattering cross section per unit mass, \( \rho _{b} \)
is the cosmic mean density, and \( r_{s} \) is the shock radius.
For all these solutions, a flat-density, isothermal core is present
which grows in size as a fixed fraction of \( r_{s} \). We find two
different regimes for these solutions: 
1) for \( Q<Q_{th}(\simeq 7.35\times 10^{-4}) \),
the core density decreases and core size increases as \( Q \) increases;
2) for \( Q>Q_{th} \),
the core density increases and core size decreases as \( Q \) increases.
Our similarity solutions are in good agreement with previous results of 
N-body simulation of SIDM halos, which correspond to the low-\( Q \) regime,
for which SIDM halo profiles match the observed galactic rotation curves
if \( Q\sim[8.4\times 10^{-4} - 4.9 \times 10^{-2}]Q_{th}\), or
\( \sigma \sim [0.56 - 5.6]\, cm^{2} g^{-1} \).
These similarity solutions also show that, as \( Q \rightarrow \infty \),
the central density acquires a singular profile, in
agreement with some earlier simulation results which approximated the effects
of SIDM collisionality by considering an ordinary fluid without conductivity,
i.e. the limit of mean free path \( \lambda_{mfp} \rightarrow 0 \). 
The intermediate regime where
\( Q\sim [18.6-231]Q_{th} \) 
or \( \sigma \sim [1.2 \times 10^{4} - 2.7 \times 10^{4}]\,cm^{2}g^{-1} \),
for which we find flat-density cores comparable to those of the low-\( Q \)
solutions preferred to make SIDM halos match halo observations, has not
previously been identified. Further study of this regime is warranted. 
}

\keywords{cosmology: dark matter --- cosmology: large-scale structure of universe --- galaxies: kinematics and dynamics --- }
\maketitle

\section{INTRODUCTION}

The cold dark matter (CDM) model provides a successful framework for 
understanding
the formation and evolution of structure in the universe. According
to this model, structure forms out of primordial density fluctuations
by hierarchical clustering; small objects form first, and then merge
to make bigger objects. However, this CDM model has several problems.

There has been a recent concern about the possible discrepancy in the 
inner density
profile of dark-matter dominated cosmological halos between N-body 
simulations and observations.
CDM halos in N-body simulations have a density 
cusp (\( \rho \propto r^{\beta } \)
where \( \beta  \) ranges from 
-1 (Navarro, Frenk \& White 1997; hereafter ``NFW'')
to -1.5 (Moore et al. 1999; hereafter ``Moore profile''),
while observations of dwarf and LSB galaxies, which are believed to
be dark-matter dominated, indicate flat-density (soft) cores.

Spergel \& Steinhardt (2000) suggested that the purely collisionless
nature of CDM be replaced by
self-interacting dark matter (SIDM) to 
resolve this discrepancy. The non-gravitational, microscopic interaction (e.g.
elastic scattering) of SIDM particles is postulated to be 
strong enough to produce a soft core
by heat conduction.
Cosmological N-body simulations which incorporate a finite scattering
cross-section  \( \sigma  \) for the SIDM particles 
show that this scheme successfully produces
soft cores in halos (e.g. Dav\'{e} et al. 2001; Yoshida et al.
2000b).

Previous analytical 
study (Balberg, Shapiro \& Inagaki 2002) and
estimates based on N-body simulations (Burkert 2000; Kochanek \& White 2000)
indicate that SIDM cores will become unstable by undergoing gravothermal
catastrophe in a Hubble time. This potential instability of SIDM cores
means that, if the matter content of the universe is really SIDM particles,
we are living in a special epoch where SIDM halos still have soft cores.
However, these analyses are based on isolated halos. In the context
of the hierarchical clustering scenario, cosmological infall is inevitable.
We are led to pose the following question:
Can cosmological infall delay the core collapse of SIDM halos?

In this paper we study the formation and evolution of SIDM halos with a
proper treatment of cosmological infall. We find that, for a range
of infall rates, the collapse (i.e. gravothermal catastrophe) of SIDM
cores can be completely inhibited. We also find a new range of values of the
scattering cross section which can produce soft cores, separate from and 
in addition to the range previously identified by N-body simulations.
Toward this end,
we find a self-similar solution to describe the evolution of SIDM
halos for arbitrary degree of collisionality which allows a detailed 
analytic study.

\section{BASIC EQUATIONS}

We use a fluid approximation derived from the Boltzmann equation
to describe the dynamics of SIDM halos. The basic
assumptions we make are as follows: 1) the system has spherical symmetry;
2) infall is radial, cold, and continuous;
3) virialized objects have an isotropic velocity distribution.
The first two assumptions have been used before for
secondary infall models (e.g. Fillmore \& Goldreich 1984). The third assumption
comes from the property of CDM halos in N-body simulations: the velocity
distribution is isotropic at the center (i.e. radial velocity dispersion
is equal to tangential velocity dispersion) and acquires only moderate
anisotropy as one goes out to the virial radius.

With these assumptions, the moments of the collisional Boltzmann equation
(i.e. valid for arbitrary degree of collisionality) can be shown to reduce
to the usual fluid equations for an ideal gas with \( \gamma =5/3 \),
where \( \gamma  \) is the ratio of specific heats.
This approach has been widely used in the literature of stellar dynamics
(e.g. Bettwieser 1983), especially for the study of gravothermal catastrophe.
We note that the particle components in both cases, that of
stellar systems (stars) and of
SIDM halos (SIDM particles), are weakly collisional, and a corresponding
conductive heating term enters the energy equation in both cases, but it has
a different origin in each case.
For the stellar systems (number of particles \( \sim \, 10^{4}-10^{6} \)),
weak, gravitational scattering is responsible for the collisions, while
for SIDM halos, nongravitational, microscopic self-interaction is responsible
for the collisions.

In summary, we use the following fluid-like equations of conservation
of mass, momentum, and energy, respectively:
\begin{equation}
\frac{\partial \rho }{\partial t}+\frac{\partial }{r^{2}\partial r}\left( r^{2}\left( \rho u\right) \right) =0,
\end{equation}
\begin{equation}
\frac{\partial }{\partial t}\left( \rho u\right) +\frac{\partial }{\partial r}\left( p+\rho u^{2}\right) +\frac{2\rho u^{2}}{r}=-\rho \frac{GM}{r^{2}},
\end{equation}
\begin{equation}
\frac{D}{Dt}\left( \frac{3p}{2\rho }\right) =-\frac{p}{\rho }\frac{\partial }{r^{2}\partial r}\left( r^{2}u\right) -\frac{\partial }{r^{2}\partial r}\left( r^{2}f\right) ,
\end{equation}
 where \( \rho  \), \( u(\equiv \left\langle v_{r} \right\rangle )\), \( p(\equiv \rho \left\langle v_{r}-u\right\rangle ^{2}) \),
and \( f \) are mass density, bulk radial velocity, pressure, and heat flux,
respectively, \(M\) is the integrated mass inside radius \(r\),
and \( \frac{D}{Dt}=\frac{\partial }{\partial t}+u\frac{\partial }{\partial r} \).
For a collisionless system, \( f=0 \) and we will call it an {}``adiabatic{}''
process. For SIDM halos, heat conduction takes the following
form:
\begin{equation}
f=-\frac{3ab\sigma }{2}\sqrt{\frac{p}{\rho }}\left( a\sigma ^{2}+\frac{4\pi G}{p}\right) ^{-1}\frac{\partial }{\partial r}\left( \frac{p}{\rho }\right) ,
\end{equation}
 where \( a \) and \( b \) are constants of 
order of unity (Balberg, Shapiro \& Inagaki 2002). We consider
an elastic scattering case, in which \( a=2.26 \) and \( b=1.002 \).

\section{SELF-SIMILAR EVOLUTION OF SIDM HALOS}

We seek a self-similar solution to describe the evolution of SIDM
halos in an Einstein-de Sitter universe. This approach makes it possible
to analyze properties of the solution in detail, because an exact
solution can be found. Previous authors have found self-similar solutions
for the evolution of cosmological halos for cold, collisionless dark matter
in radial motion (e.g. Fillmore \& Goldreich 1984) and for an ordinary,
adiabatic, baryonic fluid (e.g. Bertschinger 1985). Our problem differs from
these in that our gas is neither collisionless nor in purely radial motion
and it differs from an ordinary, adiabatic, baryonic fluid, since our
fluid-like conservation equations also contain the SIDM conduction term.
On the other hand, we were encouraged to look for a self-similar solution,
since the effect of conductivity is to drive a gas towards isothermality,
and there are self-similar equilibrium solutions known in the
limit of isothermality which have soft cores,
such as the ``truncated isothermal sphere'' (TIS)
model proposed by Shapiro, Iliev \& Raga (1999).

We start from an adiabatic (i.e. no conduction, or \( f=0 \)), self-similar
solution to find the condition required for the self-similarity of SIDM
halo solutions. If the initial overdensity has a scale-free power-law
form,
\begin{equation}
\frac{\delta M}{M}\propto M^{-\varepsilon },
\end{equation}
where \( \varepsilon  \) is any positive constant, then adiabatic
infall will form a self-similarly growing object (Fillmore \& Goldreich 1984).
A cold, supersonic, radial infall leads to a strong, spherical shock which
surrounds a subsonic region of shock-heated gas.
The corresponding length scale, e.g. a turnaround
radius \( r_{ta} \), grows as
\begin{equation}
r_{ta}\propto t^{\xi },
\end{equation}
where
\begin{equation}
\xi =\frac{2}{3}\left( \frac{3\varepsilon +1}{3\varepsilon }\right) .
\end{equation}
In this case, the shock always occurs at the same fixed fraction of
\( r_{ta} \), the mean density inside the postshock region is a constant
multiple of the time-varying cosmic mean density \( \rho_{b}\), 
and the radial profiles
of all fluid variables are self-similar.
When heat conduction by SIDM particles is introduced, however,
an additional length (time) scale is present which breaks 
the self-similarity of these
adiabatic solutions, in general.

There is one particular infall rate, however, which does not break the
self-similarity even when heat conduction is introduced.
According to the adiabatic solutions, the thermal energy changes 
according to \( \rho \frac{\partial e}{\partial t}\propto r_{ta}^{2}t^{-5} \),
while the conductive heating term would introduce a change 
as \( \frac{\partial }{r^{2}\partial r}\left( r^{2}f\right) \propto r_{ta}^{3}t^{-7} \).
To preserve similarity after the addition of heat conduction, therefore, we must
have \( r_{ta}\propto t^{2} \), or
\begin{equation}
\xi =2,\, \, \varepsilon =\frac{1}{6}.
\end{equation}

\section{SELF-SIMILAR CDM HALOS (\protect\( \varepsilon =1/6\protect \);
NO CONDUCTION)}

Let us first examine properties 
of the adiabatic solution (\( f=0 \)) for the case of \( \varepsilon =1/6 \).
We especially focus on the density profile. The density profile of the
adiabatic solution within the shock radius \( r_{s} \) agrees quite well
with the N-body results for CDM halos (see Fig. 1). 
The adiabatic solution is well-fit,
for example, by an NFW profile with concentration
parameter \( 3 \leq c_{NFW} \leq 4 \) (i.e. fractional deviation
\( \Delta \leq 17\% \) for \( 0.014 \leq r/r_{200} \leq r_{s}/r_{200} \),
where \(r_{200}\) is the NFW profile radius within which \(\left\langle\rho\right\rangle=200\rho_{b}\), and
\( r_{s} \simeq 0.6 r_{200} \)). The inner density slope of the
similarity solution
for \( 4\times 10^{-3} \leq r/r_{200} \leq 1.4 \times 10^{-2} \) is
-1.27, which is between -1 (NFW) and -1.5 (Moore profile). A natural
question is then: why does this adiabatic solution for  \( \varepsilon =1/6 \)
match the CDM N-body halos so well?
\begin{figure}[htbp]
  \begin{center}
    \leavevmode
    \epsfxsize=8.0cm
    \epsfysize=8.0cm
    \epsffile{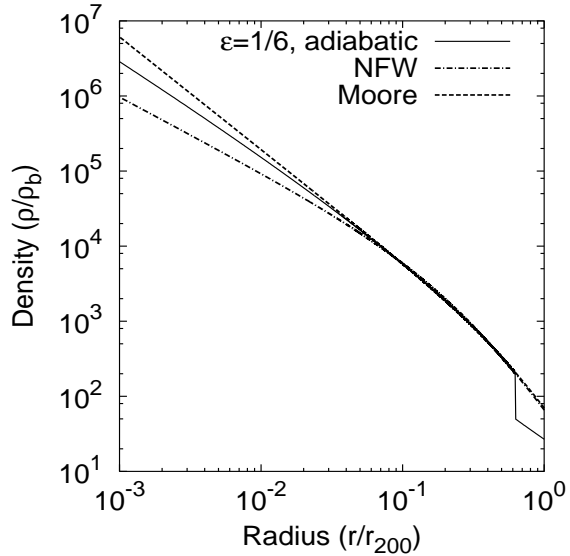}
  \end{center}
  \caption{Halo density profile for adiabatic solution
with \(\varepsilon=1/6\) for standard CDM halos, 
compared to NFW and Moore profiles. The adiabatic
solution, whose inner slope is -1.3, is a good fit to density profiles of
CDM halos from N-body simulations.}
\end{figure}

The theory of structure formation from density peaks in the Gaussian
random noise distribution of initial density fluctuations gives
an interesting clue to this correspondence.
According to Hoffman \& Shaham (1985), local maxima of Gaussian random 
fluctuations in the density can serve as the progenitors of cosmological 
structures.
For a power-law power spectrum of initial fluctuations,
\( P(k)\propto k^{n}, \) the initial density
profile can be approximated 
as \( \frac{\delta \rho }{\rho }\propto r^{\kappa }, \)
where \( \kappa=3\varepsilon=n+3 \). The value \( \varepsilon=1/6 \) corresponds
to \( \kappa=1/2, \) or \( n=-2.5 \), while \( n=-2.5 \) 
roughly corresponds to galaxy-mass
structures in CDM, as shown in Fig. 2.
\begin{figure}[htbp]
  \begin{center}
    \leavevmode
    \epsfxsize=8.0cm
    \epsfysize=8.0cm
    \epsffile{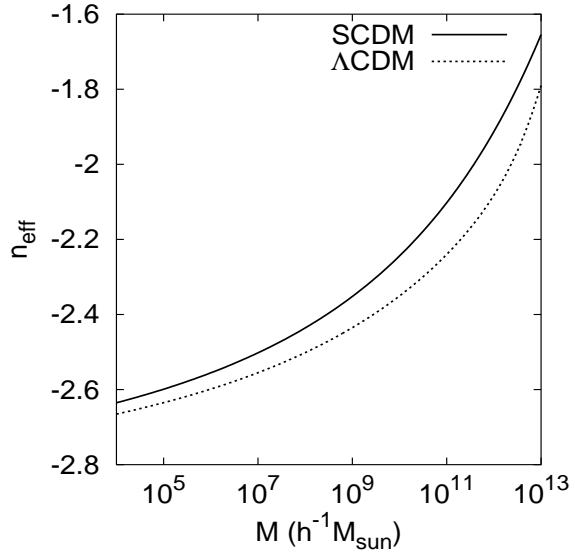}
  \end{center}
  \caption{Effective index of the power spectrum of initial density 
fluctuations (\(P(k)\propto k^{n_{eff}}\))
vs. mass of halos at their typical formation epoch, for flat, 
cluster-normalized,
matter-dominated CDM (SCDM) and COBE-normalized \(\Lambda\)CDM universes.}
\end{figure}
\begin{figure*}[t]
  \centerline{\epsfysize=13cm\epsfbox{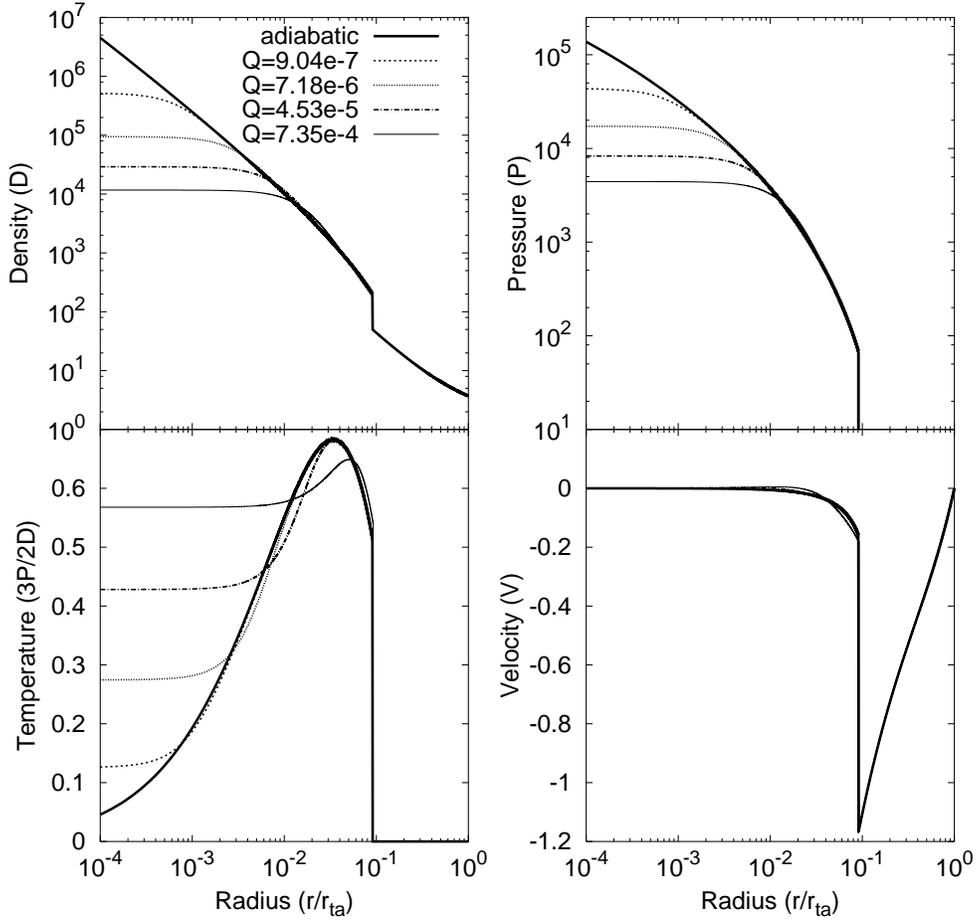}}
  \caption{Dimensionless profiles for similarity solutions in the
low-\( Q \) regime, where \(r_{ta}\) is the time-varying turnaround radius
and the shock radius is \(r_{s}=0.09r_{ta}\). Isothermal, flat-density
cores exist. As \(Q\) increases, core density decreases and core temperature
increases. 
Dimensionless similarity 
variables follow the definitions in Bertschinger (1985).}
\end{figure*}
\begin{figure*}[t]
  \centerline{\epsfysize=13cm\epsfbox{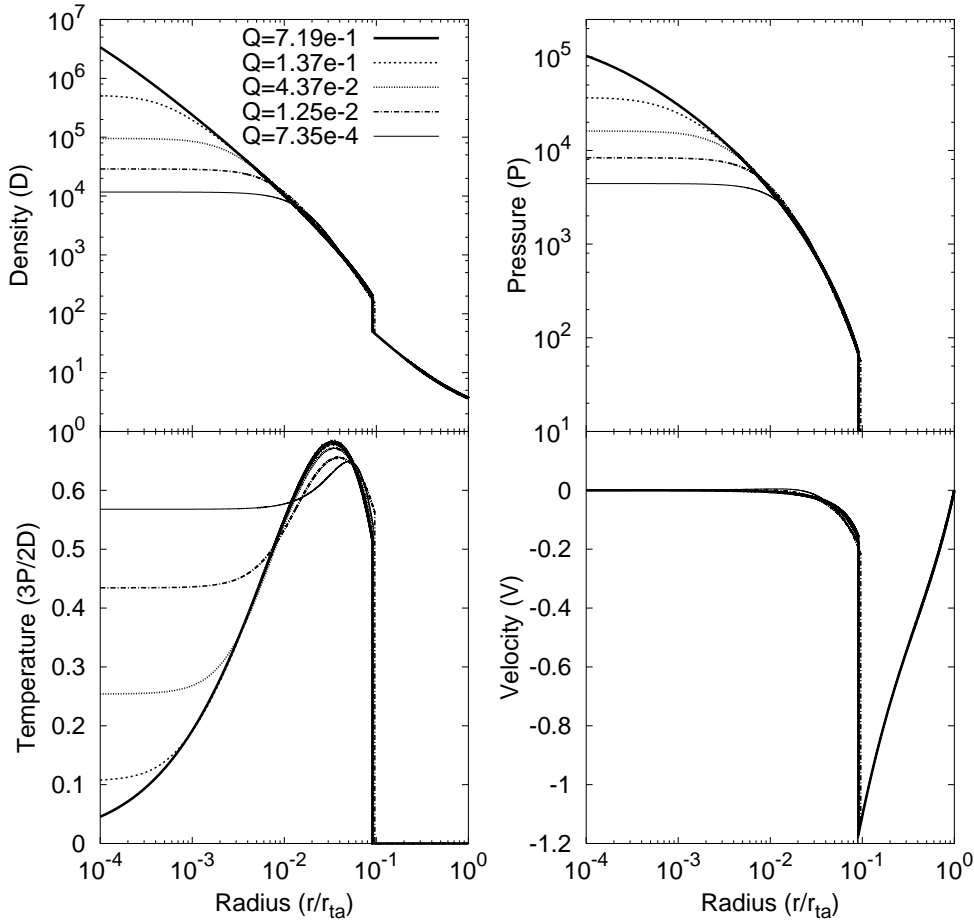}}
  \caption{Dimensionless profiles for similarity solutions in the
high-\(Q\) regime. Profiles are indistinguishable from those in Fig. 3, 
even though \(Q\) values are quite different.
The effect of \(Q\) is reversed from that of low-\(Q\) solutions: as \(Q\)
increases, core density increases and core temperature decreases.}
\end{figure*}

\section{SELF-SIMILAR SIDM HALOS (\protect\( \varepsilon =1/6\protect \);
WITH CONDUCTION)}

We find that different similarity solutions of SIDM halos arise for
different values of the collisionality parameter, 
\( Q\equiv \sigma \rho _{b}r_{s} \).
The dimensionless \(Q\) is closely related to
\( P\equiv \sigma \rho v_{rel}\Delta t \), the number of collisions
each particle experiences during a time \( \Delta t \) in a local
density \( \rho  \) with
relative velocity \( v_{rel} \), sometimes used elsewhere to parameterize the
collisionality of SIDM particles
(Dav\'{e} et al. 2001). Roughly
speaking, high \( Q \) means high collision rate. We also find that
there are two different regimes: the low-\( Q \) regime and the high-\( Q \)
regime.

In the low-\( Q \) regime, the collision mean free path in the core
is larger than the size of the halo. In this regime,
the flattening of the core density
profile increases as \( Q \) increases (see Fig. 3). As \( Q\) increases,
that is, the central density is lower and the core
radius is larger.

In the high-\( Q \) regime, the mean free path in the core region
is smaller than the size of the halo. The flattening of the core density
profile decreases as \( Q \) increases, because the mean free path
decreases (diffusion limit). As \( Q \) increases in this regime, therefore,
the core density increases and the core radius shrinks 
(see Fig. 4). The limiting case of
\( Q=\infty  \) corresponds to the solution
with no conduction, which agrees with N-body simulations with infinite
cross-section (Yoshida et al. 2000a; Moore et al. 2000). 
In this limit of maximal collisionality,
the nonadiabatic solution approaches the adiabatic solution 
(see Figs. 3 and 4).

An important aspect of this self-similar solution is that halo cores never
enter the gravothermal catastrophe phase, because cosmological infall
causes kinetic energy to be pumped continuously into the core to prevent it.
As a result, cores grow as a fixed fraction of the shock radius. 
As mentioned in Section IV, the particular value of \(\varepsilon=1/6\)
adopted here to make the evolution of SIDM halos self-similar (which,
therefore, precludes core collapse),
is natural for galaxy formation in a CDM universe, as can be
understood in the context of the growth of density peaks.
According to this theory, the average initial density 
profile responsible for the formation of individual galaxy-mass halos in
the CDM model corresponds to a value of \( n_{eff} \) which is not far from
\(-2.5\), the value required
to make \( \varepsilon=1/6 \). As such, the self-similar
behavior reported here for SIDM halos may be a realistic approximation for
galactic halos. If the infall rate which builds an individual galactic halo
eventually tapers off at late times, however, 
as reported for N-body simulations by
Wechsler et al. (2002), this will break the condition necessary
for SIDM self-similarity and may allow gravothermal catastrophe to proceed.

\section{THE ALLOWED RANGE OF SCATTERING CROSS - SECTION FOR AN SIDM UNIVERSE}

In a particular universe, we can relate \( \sigma  \) to \( Q \)
if we know the typical formation epochs for halos of different masses,
according to the Press-Schechter formalism. We find that a relatively narrow
range of Q values,  
\begin{equation}
Q\simeq[0.62 - 3.6]\times10^{-5}\left( \frac{h}{0.70}\right) \left( \frac{\sigma }{5.6cm^{2}g^{-1}}\right),
\end{equation}
characterizes the entire range of halo masses from dwarf-galaxy-mass to 
cluster-mass, in the currently-favored \(\Lambda\)CDM universe. We also find
\begin{equation}
Q\simeq[1.5 - 6.1]\times10^{-5}\left( \frac{h}{0.70}\right) \left( \frac{\sigma }{5.6cm^{2}g^{-1}}\right)
\end{equation}
for a flat, cluster-normalized, matter-dominated CDM (i.e. SCDM) 
universe (See Fig. 5).
\begin{figure}[t]
  \begin{center}
    \leavevmode
    \epsfxsize=8.0cm
    \epsfysize=8.0cm
    \epsffile{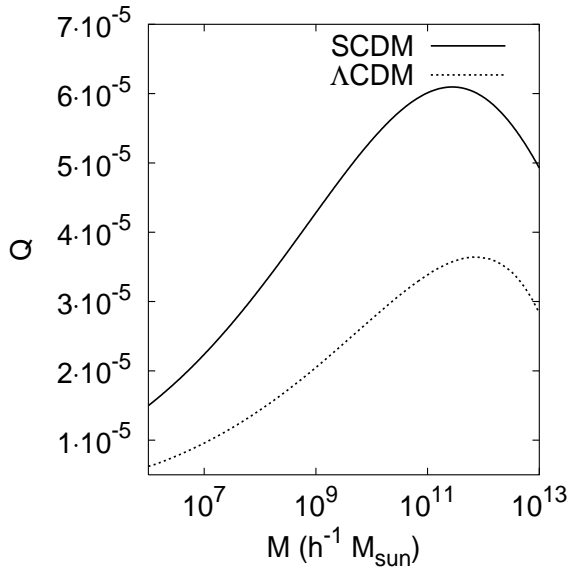}
  \end{center}
  \caption{The collisionality parameter \(Q\) vs. mass of halos
at their typical formation epoch for
\(\sigma = 5.6 \, cm^{2} g^{-1} \) and \( h=0.7 \).}
\end{figure}

The range \( \sigma =[0.56\, -\, 5.6]\,cm^{2}g^{-1} \) is the preferred range
of the scattering cross section found by cosmological N-body simulations
for the \(\Lambda\)CDM universe
to match observed galactic rotation curves (e.g. {Dav\'{e} et al. 2001}).
The above relation (equ. [9]) with \( h=0.7 \) then yields 
\( Q=[6.2\times10^{-7}\, -\, 3.6\times 10^{-5}] \) 
for the \(\Lambda\)CDM universe,
which is in the low-\( Q \) regime. 
The same profile shapes will occur in SCDM for the same Q-values.
If we apply this low-\( Q\) solution to 
SCDM with \( h=0.70 \), using equation (10), 
we get \( \sigma =[0.23\, -\, 3.3]\,cm^{2}g^{-1} \).

As shown in Section V, there also exist high-\( Q \)
solutions which yield profiles which are quite similar
to the low-\( Q \) solutions which produce observationally acceptable
soft cores. We find that \( Q=[1.37\times 10^{-2}\, -\, 1.7\times 10^{-1}] \) 
in the high-\(Q\) regime produces profiles like those in the low-\(Q\)
regime for
\( Q=[6.2\times10^{-7}\, -\, 3.6\times 10^{-5}] \).
From the relationship between 
\( \sigma  \) and \( Q \) (equs. [9] and [10]), we predict that
\( \sigma=[1.2\times 10^{4}-2.7\times 10^{4}]\,cm^{2}g^{-1} \) can also
produce acceptable soft cores, therefore, for \(\Lambda\)CDM, 
while the corresponding
\( \sigma  \) for SCDM will be about 
\( [5.11\times 10^{3}-1.56\times 10^{4}]\,cm^{2}g^{-1} \).

\section{CONCLUSION}

We have found similarity solutions which allow us to describe the dynamical
origin and evolution of SIDM halos in a fully cosmological context,
analytically, for the first time. Our solutions are based upon fluid-like
conservation equations which can be derived by taking moments of the
Boltzmann equation, assuming that the particle velocity distribution
inside virialized halos is approximately isotropic, as CDM N-body simulations
suggest. Our solutions confirm and explain the results of N-body 
simulations which have attempted to incorporate the collisional effects of SIDM
numerically within the CDM model and allow us to extend and generalize those
results to a much wider range of parameters and scales than has so far been
simulated. Along the way, we have also found a similarity solution for the
adiabatic case of standard CDM (i.e. with no SIDM self-interaction) which
serves to derive analytically the halo profile of CDM halos found previously
by N-body simulation (i.e. intermediate between NFW and Moore profiles), 
as a product of self-similar cosmological infall.

As seen in our similarity solutions, cosmological infall can affect
the dynamics of virialization significantly. 
In the context of hierarchical clustering, 
in fact,
the collapse of 
halo cores previously predicted for \(isolated\) SIDM halos is entirely
prevented by cosmological infall, according to these similarity solutions.
For realistic mass accretion histories in a CDM universe, therefore,
SIDM core collapse will be delayed until this infall becomes
negligible. Accordingly, previous analyses (Burkert 2000; Balberg, Shapiro \& Inagaki 2002)
which predict core collapse in a Hubble time should be re-examined.

By considering the full range of collisionality from \(Q=0\) to \(Q=\infty\),
we have discovered that the low-\(Q\) regime which has been found previously
by N-body simulation to produce observationally acceptable soft cores is not
unique.
We predict that a cross-section of the order of \( 10^{4}\,cm^{2}g^{-1} \)
can also produce acceptable soft cores, in addition to the regime
of smaller cross-section previously identified. Therefore, we suggest that
cosmological N-body simulations be performed which incorporate SIDM
particles with \( \sigma =[5\times 10^{3}-5\times 10^{4}]\,cm^{2}g^{-1} \)
to explore this possibility further.

In reality, the mass accretion rates of CDM halos in N-body simulations
are found to depart from the self-similar values over time, declining at
late times rather than growing (Wechsler et al. 2002).
We have accordingly begun to study the evolution
of SIDM halos with a more realistic infall, 
using a 1-D, spherical hydrodynamics code,
and will describe those results elsewhere (Ahn and Shapiro, in preparation).

\vskip 0.4cm
This work was partially supported by NASA ATP Grant NAG5-10825 and Texas
Advanced Research Program Grant 3658-0624-1999.

\end{document}